\documentclass[traditionalabstract]{aa}

\usepackage{graphicx}
\usepackage{txfonts}
\usepackage[]{hyperref} 
\usepackage{caption}
\usepackage{float}
\usepackage[]{natbib}

\usepackage{xcolor}
\hypersetup
{
    colorlinks,
    linkcolor={red!50!black},
    citecolor={blue!70!black},
    urlcolor={blue!80!black}
}

\makeatletter
\renewcommand*\aa@pageof{, page \thepage{} of \pageref*{LastPage}}
\makeatother

\usepackage{placeins}
\usepackage{subcaption}

\begin{document}

   \title{High-precision abundances of first population stars in NGC~2808: confirmation of a metallicity spread}


   \author{C.~Lardo\inst{1}\fnmsep\thanks{e-mail: carmela.lardo2@unibo.it}
          \and M.~Salaris\inst{2,3}
          \and S.~Cassisi\inst{3,4}
          \and N.~Bastian\inst{5,6}
          \and A.~Mucciarelli\inst{1,7}
          \and I.~Cabrera-Ziri\inst{8}
           \and E.~Dalessandro\inst{7}
          }

   \institute{Dipartimento di Fisica e Astronomia, Università degli Studi di Bologna, Via Gobetti 93/2, I-40129 Bologna, Italy
         \and
            Astrophysics Research Institute, Liverpool John Moores University, 146 Brownlow Hill, Liverpool L3 5RF, UK
        \and
        INAF - Osservatorio Astronomico di Abruzzo, Via M. Maggini, I-64100 Teramo, Italy
        \and
        INFN - Sezione di Pisa, Largo Pontecorvo 3, I-56127 Pisa, Italy
        \and
        Donostia International Physics Center (DIPC), Paseo Manuel de Lardizabal, 4, E-20018 Donostia-San Sebastián, Guipuzkoa, Spain
        \and
        IKERBASQUE, Basque Foundation for Science, E-48013 Bilbao, Spain
       \and
       INAF-Osservatorio di Astrofisica e Scienza dello Spazio, Via Gobetti 93/3, I-40129 Bologna, Italy
       \and
      Astronomisches Rechnen-Institut, Zentrum für Astronomie der Universität Heidelberg, Mönchhofstraße 12-14, D-69120 Heidelberg, Germany
}

   \date{Received xxx XXX, xxx; accepted xxx, XXX}

 
  \abstract{Photometric investigations have revealed that Galactic globular clusters exhibit internal metallicity variations amongst the so-called first-population stars, until now considered to have a  homogeneous initial chemical composition. This is not fully supported by the sparse spectroscopic evidence, which so far gives conflicting results.
  Here, we present a high-resolution re-analysis of five stars in the Galactic globular cluster NGC~2808 taken from the literature.
  Target stars are bright red giants with nearly identical atmospheric parameters belonging to the first population according to their identification in the chromosome map of the cluster, and we have measured precise differential abundances for Fe, Si, Ca, Ti, and Ni to the $\sim$ 0.03 dex level. 

 Thanks to the very small uncertainties associated to the differential atmospheric parameters and  abundance measurements, we find that target stars span a range of iron abundance equal to 0.25 $\pm$ 0.06~dex. The individual elemental abundances are highly correlated with the position of the star along the extended sequence described by first population objects in the 
 cluster chromosome map: bluer stars have a lower iron content. This agrees 
 with inferences from the photometric analysis.
 
 The differential abundances for all other elements also show statistically significant ranges that point to intrinsic abundance spreads. The Si, Ca, Ti, and Ni variations are highly correlated with iron variations and the total abundance spreads for all elements are consistent within the error bars. This suggests a scenario in which short-lived massive stars exploding as supernovae contributed to the self-enrichment of the gas in the natal cloud while star formation was still ongoing. 
 }


\titlerunning{Metallicity spread in NGC~2808 P1 stars}
\authorrunning{Lardo et al.}
\maketitle

\section{Introduction} \label{sec:intro}

Our views of the stellar populations hosted by Galactic globular clusters 
(GCs) have undergone a sea change during the last couple of decades. Both spectroscopic and 
photometric observations have revealed that GCs do not 
align with the standard paradigm of being populated by 
stars all with the same age and initial chemical composition; instead 
they host multiple populations (MPs) of stars that show themselves through (anti-) correlated variations 
of elements such as C, N, O, Na (in some cases also Mg and Al) and He \citep[see, e.g.,][for reviews]{gcb:12, bl18, gratton:19, csreview20}.
Stars 
with C, N, O, Na (and He) abundance patterns 
similar to those of field stars at the same [Fe/H] are usually named Population 1 (P1) --or first generation stars, according to the formation scenarios 
that envisage subsequent episodes of star formation as the origin of MPs 
\citep[see, e.g.][]{dercole08, decressin08, r22}-- while stars 
showing a range of N and Na (and He) overabundance and C and O depletion 
compared to field stars at the same [Fe/H] are named Population 2 (P2) 
or second generation stars. In the multiple star formation episodes' 
scenarios, these objects were formed later from material processed by some class of massive stars born in the first epoch of star formation.

The recent results by the {\it HST} {\sl UV legacy survey of Galactic GCs} \citep[][]{p15, milone:17} have revealed the signature of yet another chemical inhomogeneity 
among GC stars, in addition to the distinction between P1 and P2 objects.

\begin{figure}
\centering
\includegraphics[width=0.88\columnwidth]{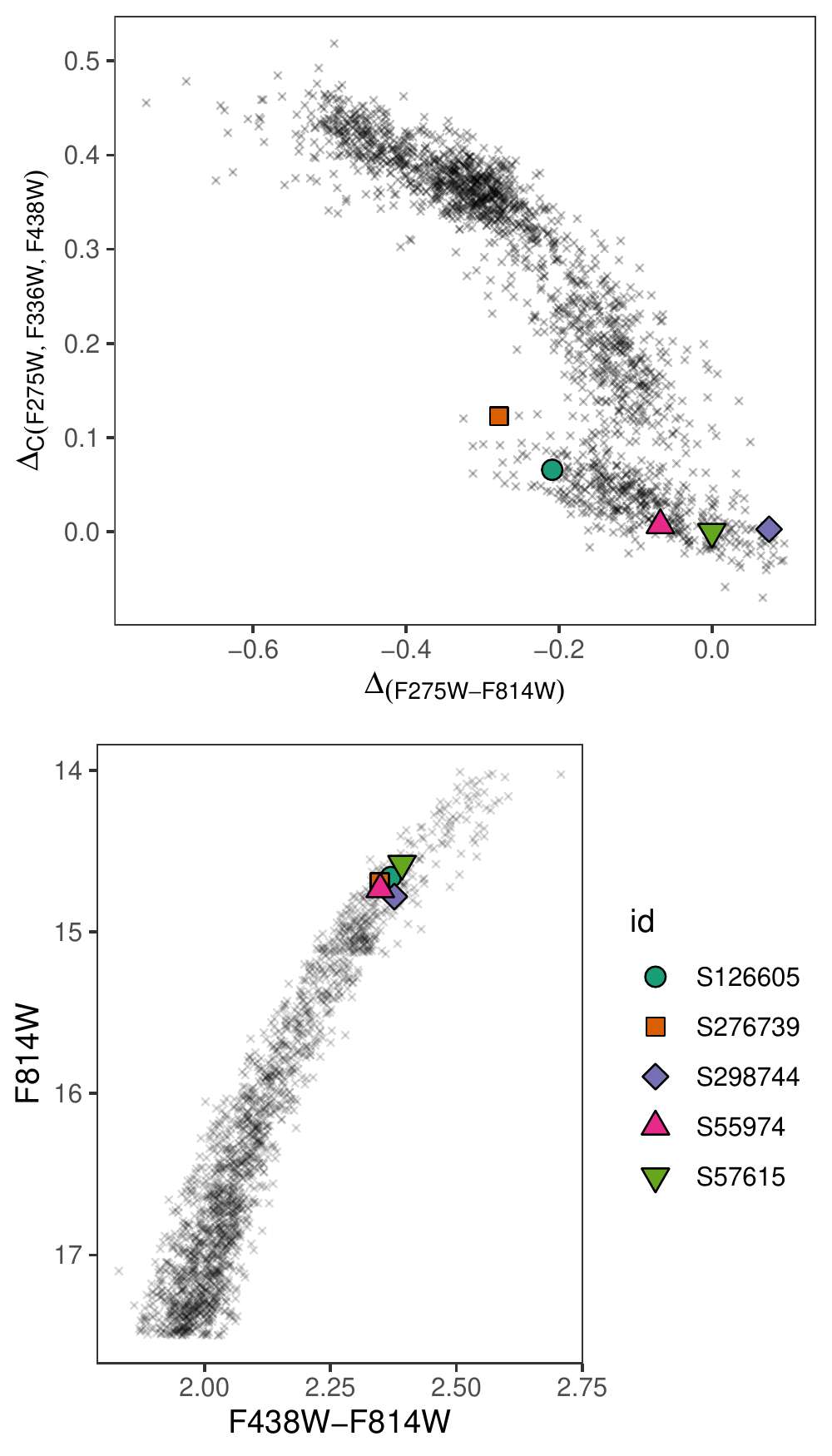}
\caption{Position of the target stars in the  chromosome map and in the $F814W-(F438W-F814W)$ CMD of NGC~2808 are shown in the top and bottom panel respectively. Photometric data are from \citet{nardiello19}. 
\label{fig:cmd}}
\end{figure}

By employing photometric filters at wavelengths shorter than 
$\sim$ 4500~\AA~that are especially sensitive to star-to-star differences 
in C, N, and O abundances \citep[see, e.g.,][]{sswc:11, p15,cassisi:13}, \citet{m17} 
have presented pseudo two-colour diagrams of red giant branch (RGB) stars  $\Delta_{\rm (F275W - F814W)}$ -  
$\Delta_{\rm C (F275W, F336W, F438W)}$
 named \lq{chromosome map\rq}
\footnote{The data employed are in the Wide Field Camera 3 filters $F275W$, $F336W$, and $F438W$  
from the {\sl UV legacy survey of Galactic GCs} \citep[see, e.g.,][]{p15}, and data in the $F814W$ filter from the Wide Field Channel of the $HST$ Advanced Camera for Survey \citep[][]{sarajedini}.} for 57 
Galactic GCs.
In these diagrams P1 and P2 stars can be easily identified, and 
a cluster's P1 RGB stars --that should be chemically homogeneous-- 
are expected to be distributed around the origin 
of the map coordinates ($\Delta_{\rm (F275W - F814W)}$$\sim$0, 
$\Delta_{\rm C (F275W,~F336W,~F438W)}$$\sim$0), spanning a 
narrow range of $\Delta_{\rm (F275W - F814W)}$  and $\Delta_{\rm C (F275W,~F336W,~F438W)}$ values. On the other hand P2 stars (with a spread of abundances of 
C, N, O, Na and He) cover a wide range of both coordinates (\citealp{milone:15, m17, carretta18}; see also Fig.~\ref{fig:cmd}). 

However, \citet{m17} have shown that the chromosome maps of the majority of their sample of Galactic GCs display spreads in their P1 sub-populations, specifically in the  $\Delta_{\rm (F275W - F814W)}$ colour. The reason for these extended P1 sequences has to be some chemical inhomogeneity among P1 stars, and variations in He and Fe have been proposed. 
Investigations by \citet{milone:15}, \citet{He18}, \citet{lsb18},
\citet{marino:19b} have demonstrated that either a range of initial He abundances 
at fixed total metallicity, or a range of metallicity at fixed He 
content, can explain the extended $\Delta_{\rm (F275W - F814W)}$ sequences, with the more metal poor or He rich P1 stars populating the lower $\Delta_{\rm (F275W - F814W)}$ values (corresponding to hotter and bluer RGB stars).

Very recently, \citet{legnardi} have investigated the two metal rich 
Galactic GCs 
NGC~6362 and NGC~6838, both showing extended P1 sequences in the chromosome maps.
They devised appropriate combinations of magnitudes 
in the $F275W$, $F336W$, $F438W$ and $F814W$ {\it HST} filters 
for P1 subgiant branch stars, able  
to disentangle the effect of metallicity and helium variations;   
by comparisons with theoretical isochrones, they found that a range of total  metallicity and not helium is present among P1 stars in these two clusters  and, by extrapolation, in all other GCs with extended P1 in their chromosome maps.

\citet{lscb} developed an alternative, independent method 
that makes use of {\it HST} near-{\sl UV} and optical photometry of RGB stars to 
disentangle the effect of metallicity and helium abundance in 
P1 stars. They applied their technique 
to the Galactic GCs M~92, NGC~2808, and NGC~6362, which cover almost 
the full range of [Fe/H] spanned by the Galactic GCs, and 
have extended P1 sequences in 
their chromosome maps, confirming that metallicity spreads are present
among their P1 stars.

These results obtained from photometric analyses imply that most of 
the Galactic GCs display a range of initial metallicities, and not just 
a handful of well known objects like $\omega$~Centauri and M~54 \citep[see, e.g.,][]{carrettametal, marino5286}.

Recent high-resolution spectroscopic investigations targeting 
specifically P1 stars have provided conflicting results. 
\citet{marino:19} have studied 18 RGB stars belonging to the extended P1 of NGC~3201, and found a range of the overall metallicity on the order of 0.1-0.15~dex. On the other hand, six RGB stars distributed along the extended P1 of the GC NGC~2808 --one of the clusters studied photometrically by \citet{lscb} -- have been investigated spectroscopically by \citet{cabrera:19}, who did not find a statistically significant spread in metallicity, at odds with the results from photometry.

Given the importance of direct spectroscopic measurements to corroborate the conclusions based on photometric methods, we present here 
a reanalysis of the chemical composition of P1 stars in NGC~2808. We made 
use of the same data published by \citet{cabrera:19} but this time instead of determining \lq{absolute\rq} abundances 
independently for all targets, 
we have performed a purely differential analysis, measuring the relative abundances of several metals with respect to a reference star. This way, the effect of systematic errors that add substantially to the total error budget on the chemical abundances are minimized, and small metallicity differences can  be revealed with much higher statistical significance.
Section~\ref{data} presents briefly the spectroscopic data, followed in Sect.~\ref{analysis} by a description of our analysis, and by a discussion of the results in Sect.~\ref{discussion}. Our conclusions are presented in Sect.\ref{conclusions}.


\section{Data}\label{data}
The five P1 stars analysed in this work are from \citet{cabrera:19}\footnote{We do not include in our sample star S187128, which belongs to the P2 sub-population according to its element abundance pattern \citep{cabrera:19}.}.
Targets are placed in an extremely narrow region of the $F814W-(F438W-F814W)$ colour magnitude diagram (CMD; see Figure~\ref{fig:cmd}), with similar optical colours and magnitudes. This minimises the impact on the spectra of differences in stellar atmospheric parameters. However, they cover the full $\Delta_{\rm (F275W - F814W)}$ extension of the P1 population in the chromosome map of NGC~2808, as shown in Figure~\ref{fig:cmd}. Also, they are all confirmed cluster members according to proper motions and radial velocities \citep{cabrera:19}.

High-resolution spectra were taken with MIKE at the Magellan-Clay telescope \citep{bernstein03} using the 0.7 $\times$ 5 arcsec$^2$ slit, which provides a spectral resolution of $\sim$40\ 000 in the red arm. Raw spectra were reduced with the {\tt CarPy} version of the pipeline \citep{kelson:00,kelson:03}. The typical signal-to-noise ratio of MIKE spectra is $\approx$50-60 at 5800\AA. Additional details can be found in \citet{cabrera:19}.

\begin{figure*}
\centering
\includegraphics[width=0.82\textwidth]{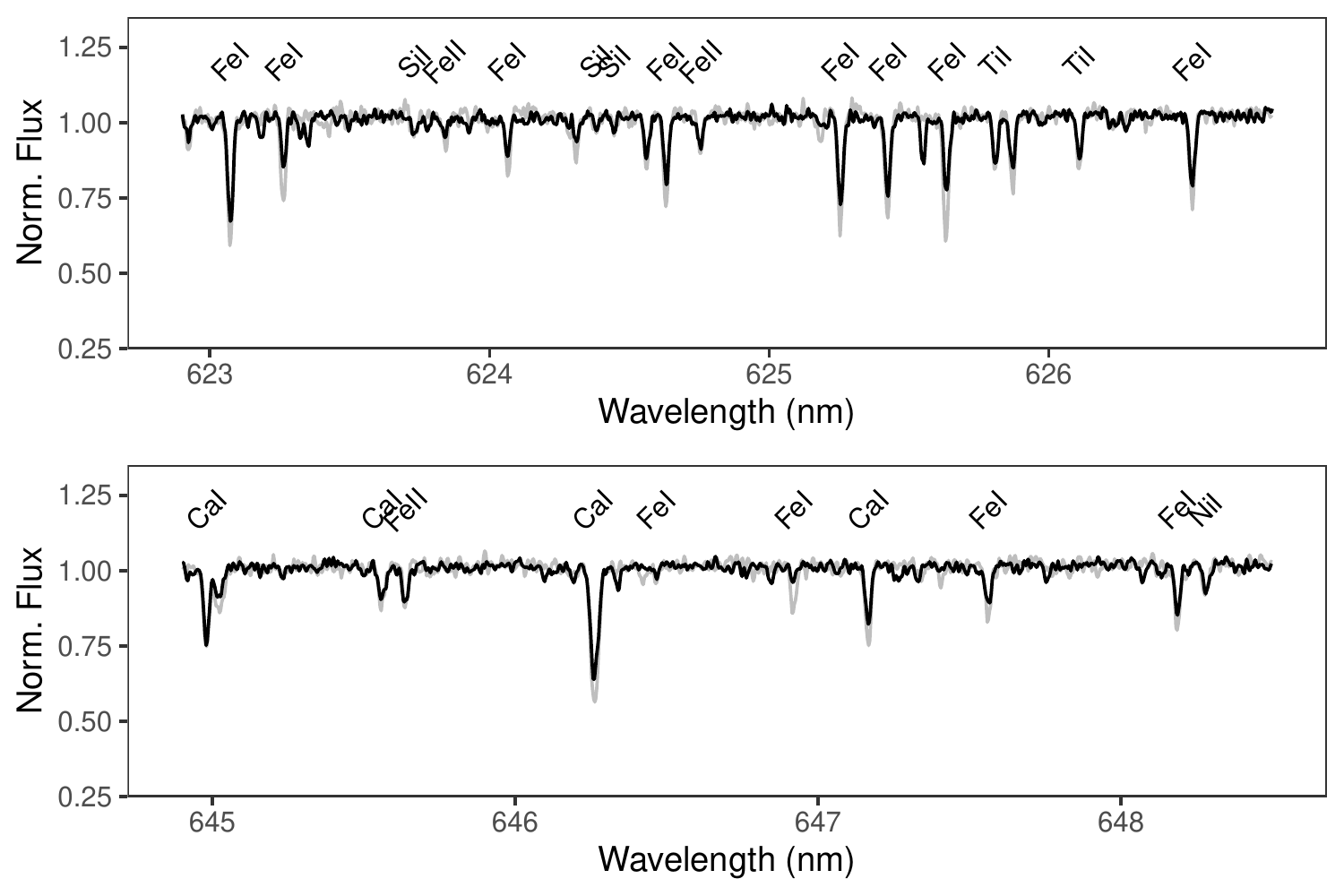}
\caption{Spectra of stars S276739 (black) and S57615 (light grey) with very similar positions in the optical CMD. Some prominent metallic lines are labelled.
\label{fig:spectra}}
\end{figure*}

Figure~\ref{fig:spectra} shows an example of the reduced one dimensional spectra for stars S276739 and S57615, in two wavelength regions. These stars occupy essentially the same place in the optical CMD of the cluster, but two 
very different locations along the P1 sequence 
in the chromosome map (see Fig.~\ref{fig:cmd}) and show very different strengths of the metallic lines. This suggests the need for reanalysing \citet{cabrera:19} stars using a differential technique able to attain a very high precision in the determination of relative abundances.


\section{Chemical analysis}\label{analysis}
Given that we are interested in establishing whether abundance variations are present amongst P1 stars, we have performed a differential line-by-line analysis with respect to a selected reference star. This has allowed us to achieve a precision in the measurement of the relative abundances 
equal to $\sim$0.03~dex  \citep[e.g.;][]{melendez09,melendez12,alves10, ramirez12,ramirez14,yong13,reggiani16,spina18,nissen18,casamiquela20,mckenzie22}. 

We have made use of the software  {\tt qoyllur-quipu}  ({\tt q$^2$}) \citep{ramirez14}\footnote{{\tt q$^2$} is a Python package freely available at \url{https://github.com/astroChasqui/q2}.}, which employs the 2019 version of the spectrum synthesis code {\tt MOOG} \citep{moog} for the calculations (specifically the {\tt abfind} driver). We adopted the standard metal distribution of the MARCS grid of 1D-LTE model atmospheres \citep{marcs}, and interpolated the model atmospheres linearly to the input atmospheric parameter values when necessary.  

The equivalent widths (EWs) were measured with the code {\tt DAOSPEC} \citep{daospec} through the wrapper {\tt 4DAO} \citep{mucc:13,muc:17}. The atomic linelist is from \citet{heiter:21}.  Only lines with a EW between 10 and 120 m\AA~were considered in the abundance analysis, to avoid weak and noisy lines, as well as very strong features in the flat part of the curve of growth. Moreover, we have considered only lines between 4800 and 6800\AA, to sample the region with the highest signal-to-noise ratio not affected by telluric absorption. 
All the lines with EW uncertainties larger than 15\%\footnote{Uncertainties in the EW measurements are estimated by {\tt DAOSPEC} as the standard deviation of the local flux residuals, and represent a 68\% confidence interval of the derived EW.} are also excluded. Finally, to compute the abundance of iron, we kept only lines within 1$\sigma$ from the median iron value.


\begin{figure}
\centering
\includegraphics[width=0.6\columnwidth]{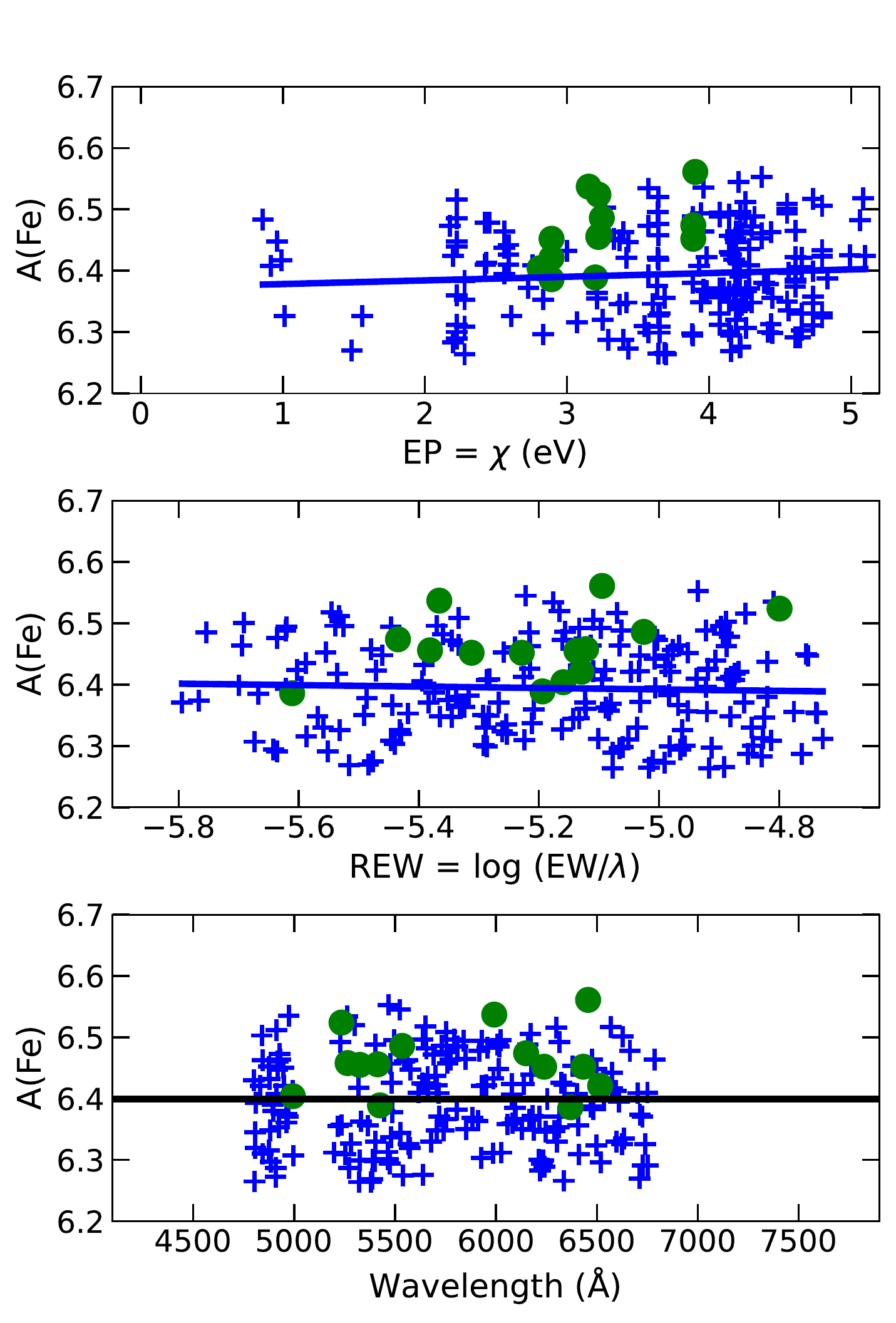}
\caption{The absolute abundances from individual neutral iron lines A(Fe~I) for the reference object S55974 are plotted against 
their excitation potential (EP), reduced equivalent width (REW) and  wavelength (from top to bottom) using the atmospheric parameters from \citet{cabrera:19}. Blue crosses denote abundances from the Fe I lines and green circles abundances from the Fe II lines. The solid blue lines are linear fits to the Fe~I  data and the black line in the bottom panel is a horizontal line at the average value of the iron abundance. 
\label{fig:par_photo}}
\end{figure}


\begin{figure}
\centering
\includegraphics[width=0.6\columnwidth]{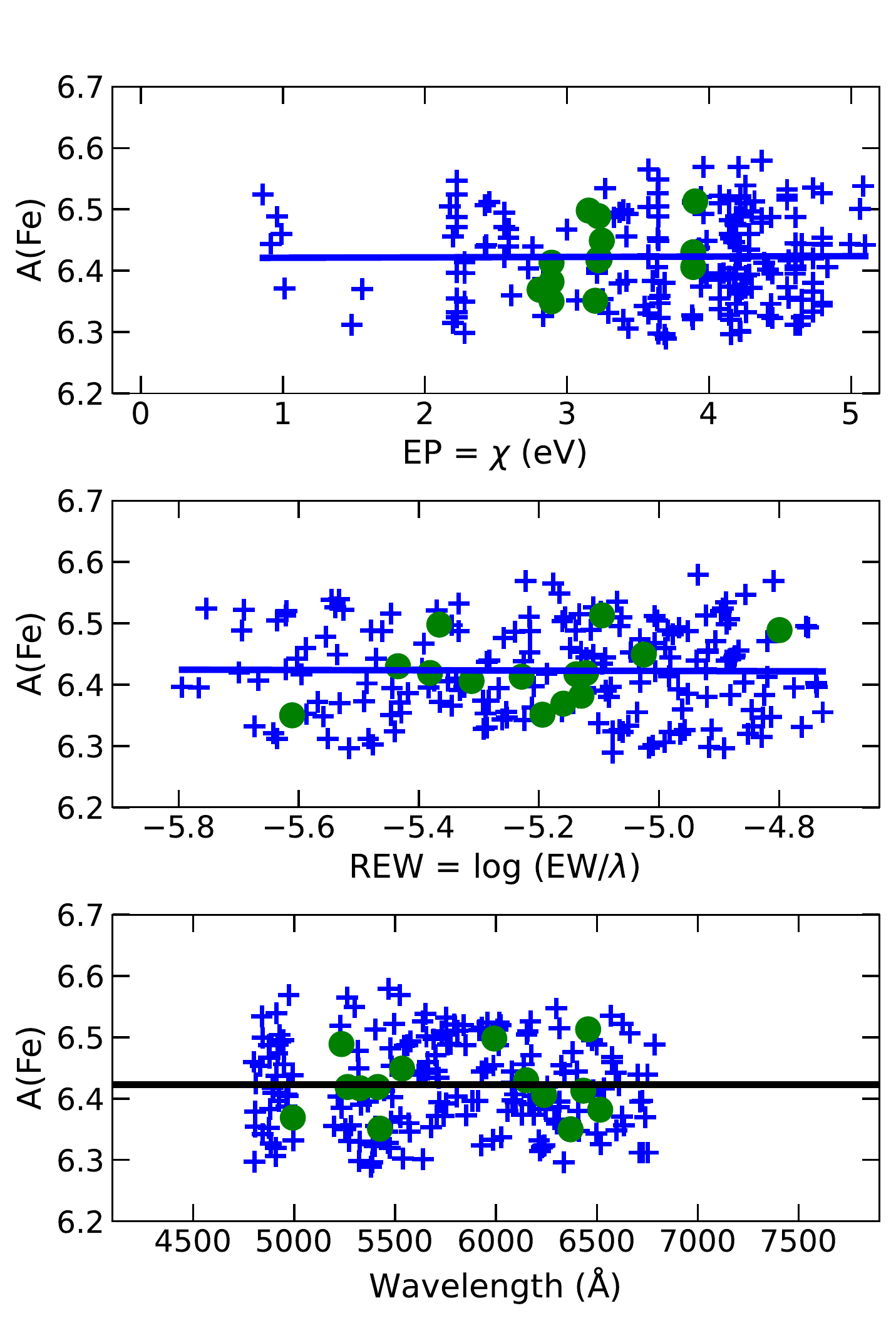}
\caption{The same as Figure~\ref{fig:par_photo} but for the atmospheric parameters computed using the standard spectroscopic approach. In particular, in this case we are using iron ionising equilibrium as gravity indicator.} 
\label{fig:par_spectro}
\end{figure}

\subsection{Atmospheric parameters for the reference star}
We have selected as reference object the star S55974,  because its effective temperature (T$_{\rm eff}$) is close to the median value of the sample stars \citep{cabrera:19}.

Initial guesses for effective temperature (T$_{\rm eff}$), surface gravity ($\log$g), micro-turbulence velocity ($v_{\rm t}$), and metallicity ([Fe/H]) are from \citet{cabrera:19}. 
Specifically, these authors have derived T$_{\rm eff}$ from the dereddened $(V-I)$ colour \citep{saraj:07} using the \citep{alo99,alo01} calibration. Surface gravities were estimated using the computed T$_{\rm eff}$ and the stellar luminosity, as determined from the photometry by assuming a RGB mass equal to 0.8 M$_{\odot}$, a reddening $E(B-V)$ = 0.22 and a distance modulus $(m-M)_{\rm V}$= 15.59~\citep{harris}, and bolometric corrections from \citet{alo99}. 
Finally, microturbulent velocities were obtained by erasing any trend between the mean iron abundance derived from the  Fe~I lines and the logarithm of the reduced equivalent widths,  defined as $\log ({\rm REW}) = \log ({\rm EW}/\lambda)$, where EW is the equivalent width of the line centred at wavelength $\lambda$. 

Figure~\ref{fig:par_photo} shows the diagnostic plots for S55974 when the photometric parameters from \citet{cabrera:19} are adopted. In particular, we set T$_{\rm eff}$ = 4809 K, $\log$(g) = 2.15 dex, v$_t$ = 1.30 km/s and derived for S55974 absolute iron abundances  A(Fe I)  = 6.39 $\pm$ 0.07, and 
A(Fe II) = 6.46 $\pm$ 0.05 for the neutral and ionised iron lines, respectively.
The fact that the iron abundance shows no trend with the excitation potential (EP) and the reduced equivalent width (REW), suggests that both the effective temperature and microturbulence are neither  underestimated or overestimated. Iron abundances from Fe~I and Fe~II are different, since we were not imposing ionising equilibrium to infer gravities, but the derived abundances are fully consistent within the errors. This hints that surface gravity is also neither underestimated or overestimated. 
The slopes of the linear fit between absolute iron abundance A(Fe~I ) and EP and REW are in both cases zero within the errors (A(Fe I) vs. EP has a slope equal to 0.006 $\pm$ 0.006, and A(Fe I) vs. REW has a slope equal to --0.012 $\pm$ 0.021). The same applies to absolute iron abundance from  ionised iron absorption lines.

Starting from those initial T$_{\rm eff}$, $\log g$, and $v_t$, spectroscopic stellar parameters for star S55974 have been computed using the standard excitation/ionisation balance technique. In particular, we use {\tt q$^2$} to find those values that minimise the slopes of iron abundance versus the line EP and REW (e.g. to derive T$_{\rm eff}$ and $v_t$; respectively), as well as match the average Fe I and Fe II abundances (e.g., to estimate $\log g$), iterating by varying T$_{\rm eff}$, $\log g$, and $v_t$ in steps of 8~K, 0.08~dex, and 0.08 km/s, respectively, until {\tt q$^2$} converges.


\begin{figure*}

\centering
  \begin{subfigure}{0.35\textwidth}
     \centering\includegraphics[width=0.85\textwidth]{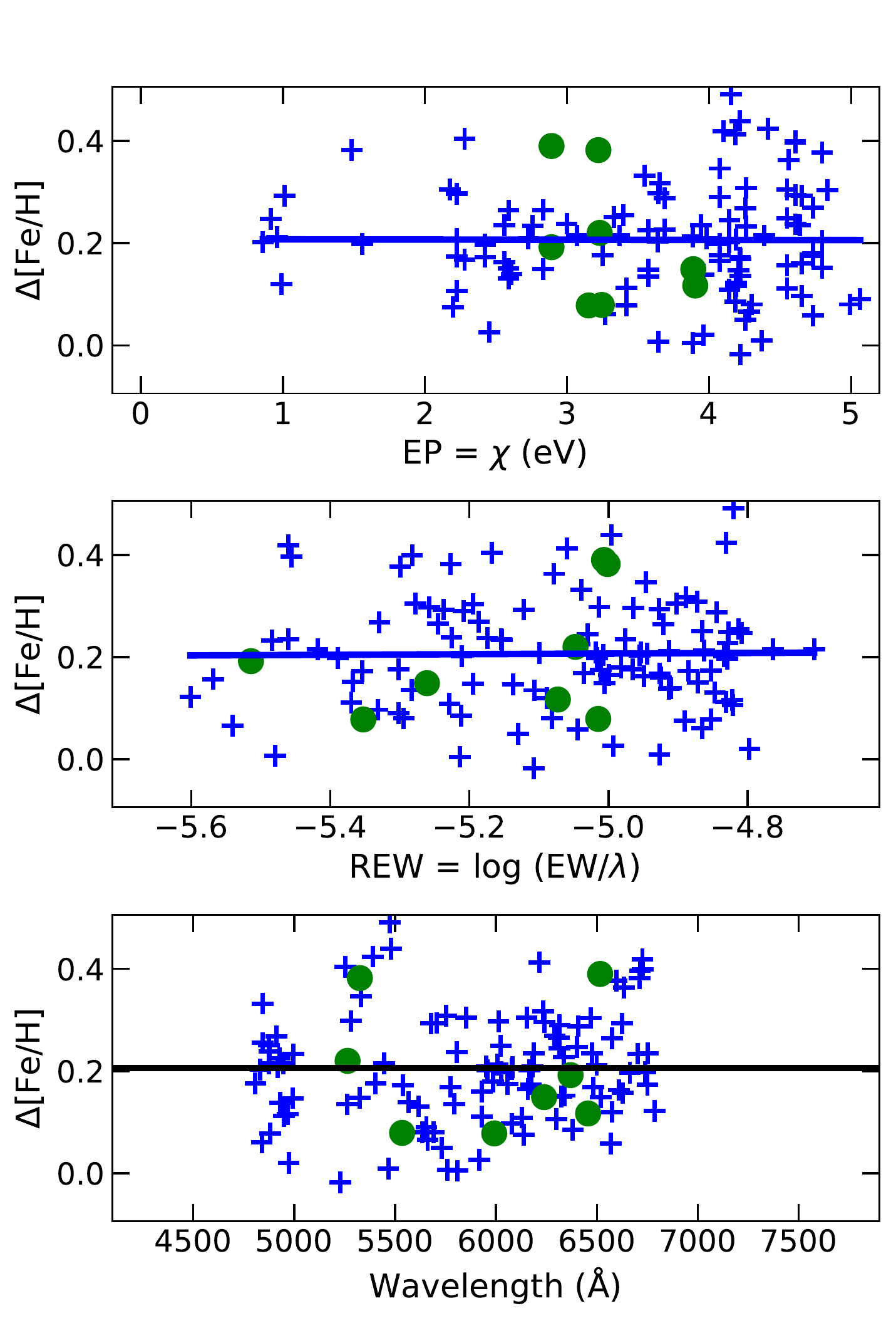}
    \caption{S57615}
  \end{subfigure}
 \centering
  \begin{subfigure}{0.35\textwidth}
    \centering
    \includegraphics[width=0.85\textwidth]{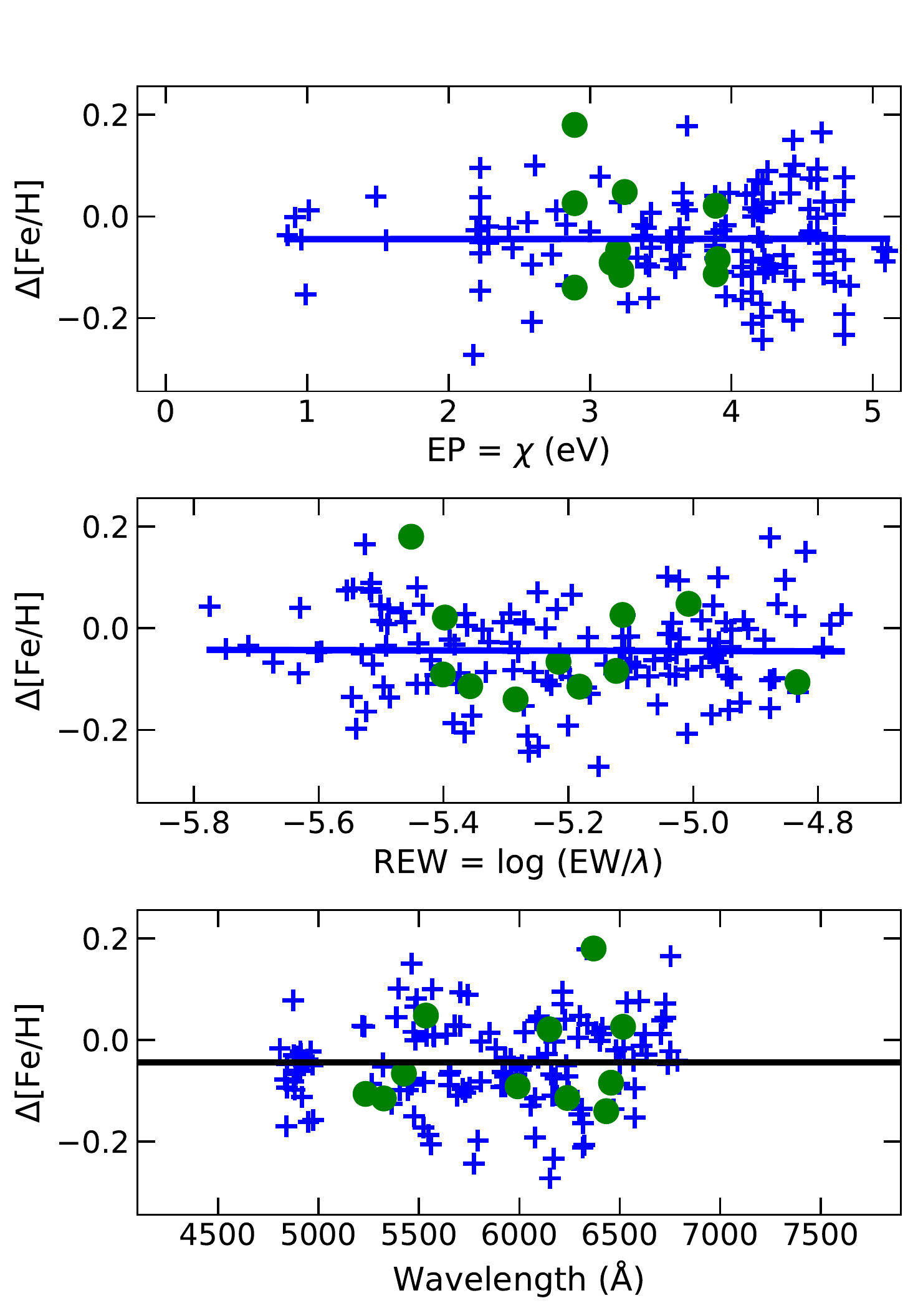}
    \caption{S126605}
  \end{subfigure}
  \centering
  \begin{subfigure}{0.35\textwidth}
    \centering\includegraphics[width=0.85\textwidth]{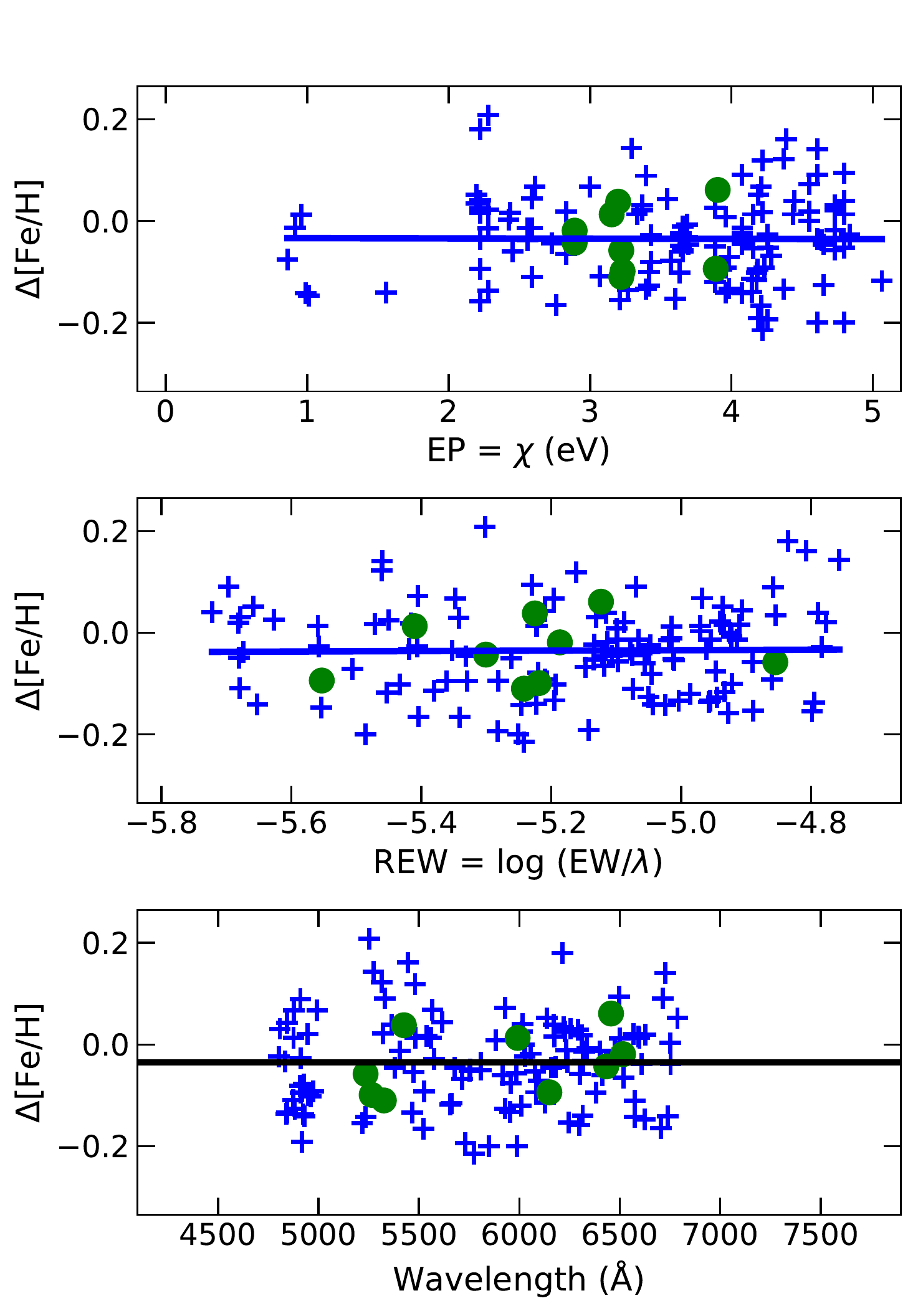}
    \caption{S276739}
  \end{subfigure}
  \centering
  \begin{subfigure}{0.35\textwidth}
    \centering\includegraphics[width=0.85\textwidth]{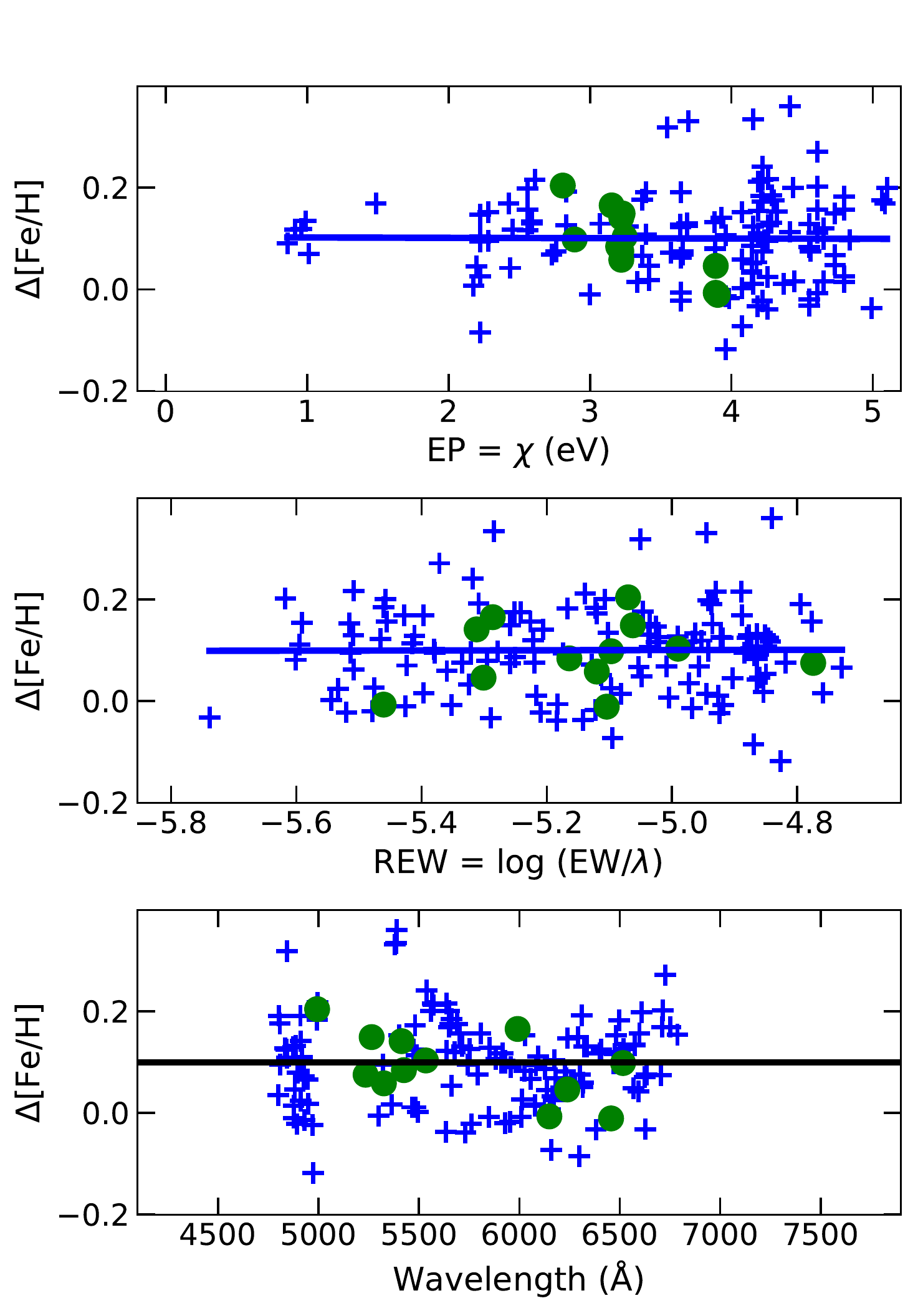}
    \caption{S298744}
  \end{subfigure}
  \vfill
  
  \caption{Differential line-by-line iron abundances for S57615, S126605, S276739, and S298744 (from top-left to bottom-right) are shown as a function of EP, REW, and wavelength. Each symbol denotes the differential (neutral and ionised) iron abundance with respect to the reference star S55974.  Symbols are the same as in Figure~\ref{fig:par_photo}.
\label{fig:diagnostic}}
\end{figure*}

After optimisation, we derive atmospheric parameters that are fully compatible with the ones inferred from photometry by \citet{cabrera:19}. We find T$_{\rm eff}$ = 4834 K, $\log$(g) = 2.06 dex, and v$_t$ = 1.29 km/s, with formal 1$\sigma$ errors 
equal to eT$_{\rm eff}$ = 26 K, e$\log$g = 0.07 dex, and ev$_t$ = 0.04 km/s respectively. These values represent the precision at minimising Fe abundance trends and Fe I vs. Fe II iron abundance differences. Thus they do not reflect the full uncertainties in stellar parameters, which are dominated by systematic uncertainties. 

Diagnostic plots for the spectroscopic parameter case are in Figure~\ref{fig:par_spectro}. In this case, the derived absolute iron abundances for neutral and ionised iron lines are of A(Fe I)  = 6.42 $\pm$ 0.07, and A(Fe II) = 6.42 $\pm$ 0.05; respectively. Thus, absolute iron abundances A(Fe I) for the reference star S55974 differ by only +0.03 dex when spectroscopic parameters are adopted instead of those based on photometry.  
Also in this case, the slopes of the linear fit between A(Fe I) vs. EP, and A(Fe I) vs REW are virtually zero (A(Fe I) vs. EP slope  =  0.001 $\pm$ 0.006, and A(Fe I) vs. REW slope = --0.003 $\pm$ 0.02).

\subsection{Line-by-line differential stellar parameters}\label{par_dis}

Once we have determined T$_{\rm eff}$, $\log g$, and $v_t$, and iron abundances for the reference star S55974, we moved to the differential analysis for the other stars in the sample \citep[see][where high precision differential abundance measurements are obtained for NGC~6752 and M~22; respectively]{yong13,mckenzie22}.

The differences  $\Delta$[Fe/H] for each object were measured relative to the iron abundance of the reference  star S55974 on a line-by-line basis. If [Fe/H]$_{\rm i}$ is the iron abundance derived for a given iron line i , the abundance difference (program star - reference star) for the same line is:  
$$ \Delta {\rm [Fe/H]}_{\rm i}=  {\rm [Fe/H]}_{\rm i}^{\rm star} - {\rm [Fe/H]}_{\rm i}^{\rm reference}.$$

We then proceed with the analysis using the standard spectroscopic approach.
First, we applied the condition of excitation equilibrium by minimising the slopes of these abundance differences for Fe I vs. EP; thus imposing the following constraint:

$$ \frac{\partial (\Delta {\rm [Fe/H]}_{\rm i})}{\partial {\rm (EP)}}=0.$$

Secondly,  we considered the abundance differences for Fe I as a function of reduced equivalent width, REW, and imposed the following constraint:

$$ \frac{\partial (\Delta {\rm [Fe/H]}_{\rm i})}{\partial {\rm (REW)}}=0.$$

This allowed us to minimise the impact of model uncertainties as well as errors in the atomic data because they cancel out in each line calculation. This is particularly true in our case, given that all stars have very similar temperatures and gravities and they are also similar to the star adopted as reference. 

We then defined the average abundance difference for iron as:

$$\Delta {\rm [Fe/H]} = \frac{1}{\rm N} \sum_{i=1}^{N}  \Delta {\rm [Fe/H]}_{\rm i}.$$

where N is the number of lines considered.

Figure~\ref{fig:diagnostic} shows the diagnostic plots (e.g., abundance versus EP/REW/wavelength plots) for the programme stars used to derive atmospheric parameters and iron abundances with respect to the reference star S55974. Note that each symbol denotes the differential (neutral and ionised) iron abundance with respect to the reference star S55974. In all cases, the slopes of the linear fit between $\Delta {\rm [Fe/H]}$ vs. EP, and $\Delta {\rm [Fe/H]}$ vs REW are zero.

 \begin{table*}
 \centering
 \begin{footnotesize}
  \setlength{\tabcolsep}{3pt}    
\caption{Atmospheric parameters and differential abundances with respect to star S55974.
The last line lists the absolute abundances for the reference star S55974.
The number of lines considered to derive differential abundances is reported in parentheses.}              
\label{table:par}      
\centering                                  
\begin{tabular}{l l l l r  r  r  r  r }        
\hline\hline 
\footnotesize
 
Id & T$_{\rm eff}$ & $\log$(g) & v$_t$ & $\Delta$[Fe/H]   & $\Delta$[Si/H]   & $\Delta$[Ca/H]   & $\Delta$[Ti II/H]   & $\Delta$[Ni/H]  \\
\hline                                 
S57615  & 4841 $\pm$ 42& 2.05 $\pm$ 0.11&  1.35$\pm$0.08  &   0.21$\pm$0.05  (116)  &  0.19$\pm$	0.06 	(5) &	0.16$\pm$	0.07 (4) &	0.25$\pm$	0.07 (3) &	0.19$\pm$	0.06  	(8)  \\
S126605 & 4868 $\pm$ 20& 1.94 $\pm$ 0.10&  1.14$\pm$0.04  &   --0.04$\pm$0.02  (143)  & -0.08$\pm$	0.04 	(7) &	-0.03 $\pm$	0.03 (5) &	-0.14$\pm$	0.07 (4) &	-0.10$\pm$	0.04 (7)  \\

S276739 & 4838 $\pm$ 32& 2.14 $\pm$0.09&  1.09$\pm$0.05  &   --0.04$\pm$0.03  (135)  & 0.04$\pm$	0.04 	(5)  &	-0.06 $\pm$	0.05 (7) &	-0.08$\pm$	0.07 (4) &	-0.02$\pm$	0.05 (7) \\
S298744 & 4877 $\pm$ 32& 2.10 $\pm$ 0.09&  1.44$\pm$0.06  &   0.10$\pm$0.03   (144)  & 0.17$\pm$	0.03	(6) &	0.02 $\pm$	0.05  (4) &	0.11$\pm$	0.07 (4) &	0.11$\pm$	0.05 (8) \\
\hline
 & & & & & & & &  \\
\hline
\hline
Id & T$_{\rm eff}$ & $\log$(g) & v$_t$ & [Fe/H]   & [Si/H]   & [Ca/H]   & [Ti II/H]   & [Ni/H]  \\
\hline
S55974 & 4834 $\pm$ 26& 2.06 $\pm$ 0.07 & 1.29 $\pm$ 0.04 &   --1.03 $\pm$ 0.07 (196) & 
--0.92 $\pm$  0.07 (7) & --0.66 $\pm$  0.03 (7) & --0.88 $\pm$  0.08 (5) &  --1.12 $\pm$  0.08 (11) \\

\hline                                            
\end{tabular}
\end{footnotesize}
\end{table*}


\begin{figure}
\centering
\includegraphics[width=0.7\columnwidth]{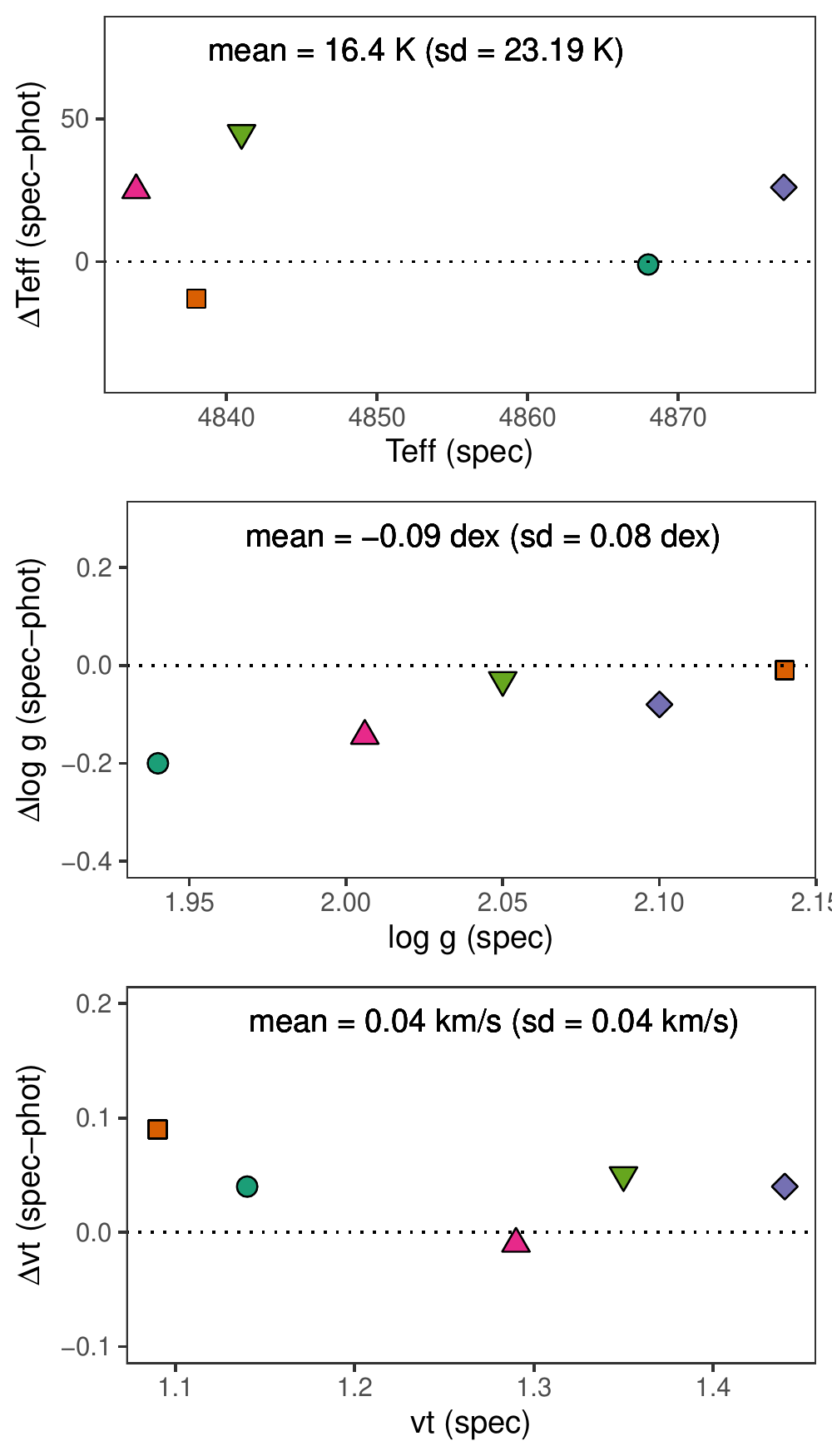}
\caption{Difference between the values determined by  \citet{cabrera:19} from photometry, and their counterpart from spectroscopy, for effective temperature ($\Delta$ T$_{\rm eff}$), surface gravity ($\Delta$ $\log$ g) and microturbolent velocity ($\Delta$ v$_{\rm t}$),  respectively. Symbols are the same as in Figure~\ref{fig:cmd}. The average differences with respect to \citet{cabrera:19}, along with their associated standard deviation, are listed in each panel.
\label{fig:par_atm}}
\end{figure}

Next, we present a comparison between our final values of the atmospheric parameters  T$_{\rm eff}$, $\log$ g and  v$_{\rm t}$ 
and those derived by \citet{cabrera:19} from photometry in Fig.~\ref{fig:par_atm}.
This figure shows that the agreement between the two studies is excellent: differences are vanishingly small and in all cases within the typical errors  quoted in \citet{cabrera:19}, amounting for T$_{\rm eff}$, $\log$ g and  v$_{\rm t}$ to $\pm$75~K, $\pm$0.2 dex and $\pm$0.2 km/s, respectively \citep[see][for a discussion on the applicability of the spectroscopic approach to GC studies across the entire metallicity range]{mucciarelli:20}.

The final parameters and differential iron abundances are listed in Table~\ref{table:par}. This table also reports the formal errors associated to the measurements, which  represent the internal precision of the technique  \citep[e.g.;][]{epstein,bensby}.
The average internal errors on the derived atmospheric parameters and differential abundances are: $\sigma$ (T$_{\rm eff}$)= 32K, $\sigma$ ($\log$ g)  = 0.10 dex, $\sigma (v_{\rm t})$ = 0.06 km/s, and  $\sigma (\Delta$[Fe/H]) =0.03 dex. 
They do not reflect the true uncertainties, that are dominated by systematic errors, however this is not an issue for our analysis because we are interested in measuring abundance differences rather than absolute values.

\subsection{Line-by-line differential chemical abundances}

Differential abundances have been also calculated for all the elements with more than three clean and relatively strong features in the wavelength range considered. 
For any species (X), we define the average abundance difference in a manner similar to what we did for iron:

$$\Delta {\rm [X/H]} = \frac{1}{\rm N} \sum_{i=1}^{N}  \Delta {\rm [X/H]}_{\rm i},$$

where:

$$ \Delta {\rm [X/H]}_{\rm i}=  {\rm [X/H]}_{\rm i}^{\rm star} - {\rm [X/H]}_{\rm i}^{\rm reference};$$

with [X/H]$_{\rm i}$ the elemental abundance derived for a given absorption line i. 
We have accounted for lines of Si, Ca,  Ti~II, and Ni, and the derived abundance differences are listed in Table~\ref{table:par}. The same table also gives the error associated to the differential abundance measurements. As seen in Table~\ref{table:par}, the 
range of relative abundances 
for all the measured elements are substantially larger than the individual measurement uncertainties.  

Finally, we did not consider in our analysis non-local thermodynamic equilibrium (NLTE) effects. Indeed, the impact of NLTE corrections on relative abundances is negligible because stars occupy the same region in the optical CMD of the cluster \citep{lind12}.

\section{Discussion}
\label{discussion}
Because of the small errors associated to our derived differential abundances, our analysis is capable of revealing subtle differences in the abundances of P1 stars. 
Indeed, Table~\ref{table:par} shows 
that the range of relative 
abundances for all elements is much larger than the average uncertainty on the individual values.
In the case of Fe, for example, we find a range equal to 0.25 $\pm$ 0.06 dex, i.e. a 
Fe abundance spread among our sample of P1 
stars that is significant at more than 
the 3$\sigma$ level.

The individual absolute abundances determined 
by \citet{cabrera:19} provided a 
spread equal 0.23$\pm$ 0.13~dex, similar 
to our result but with a much larger error, 
which implied a low statistical significance and led the authors to conclude that there is no 
metallicity spread among P1 stars in this cluster. We also note that the lack of any detection of intrinsic iron variations in \citet{cabrera:19} is not due to the different set of atmospheric parameters used in the two studies (e.g., the differences are extremely small and in any case within the uncertainties associated to the photometric determination; see~\ref{par_dis}), but rather to the much smaller formal errors associated to our differential abundances. Thus, high-precision relative abundances confirm the presence of star-to-star metallicity variations in NGC2808 P1 stars as found by photometric studies \citep{legnardi,lscb}. However, the spread inferred from the analysis by \citet{legnardi} ($\Delta$[Fe/H]$_{\rm P1}$= 0.11$\pm$0.11) is about half the value we derived from spectroscopy. 


\begin{figure}
\centering
\includegraphics[width=0.6\columnwidth]{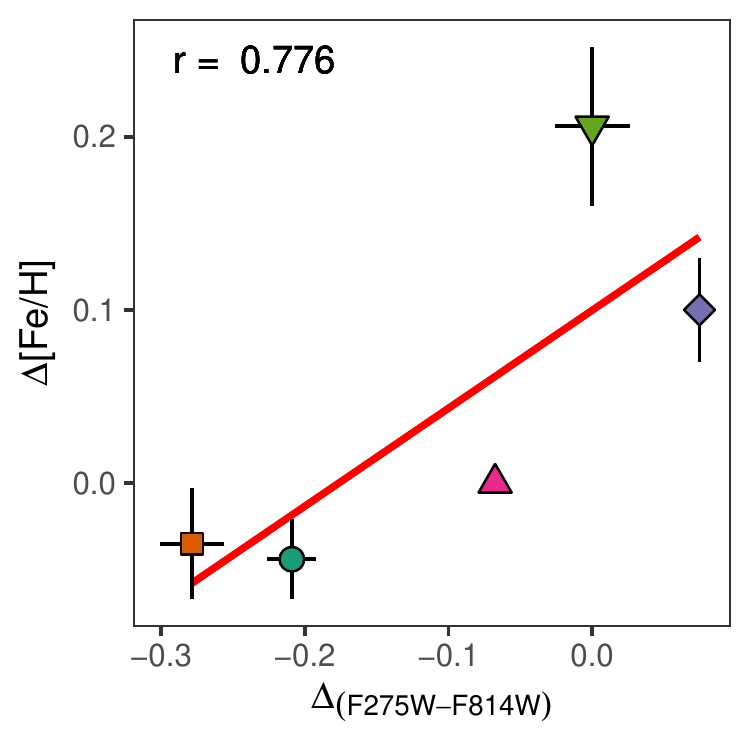}
\caption{Abundance differences $\Delta$[Fe/H] with respect to the reference star S55974 (also swown in the plot), as a function of the corresponding $\Delta _{\rm F275W,F814W}$ coordinate in the chromosome map of NGC~2808. The solid red line is a linear fit to the data. The Pearson correlation coefficient is reported in the top left corner. 
\label{fig:iron_spread}}
\end{figure}


\begin{figure}
\centering
\includegraphics[width=1\columnwidth]{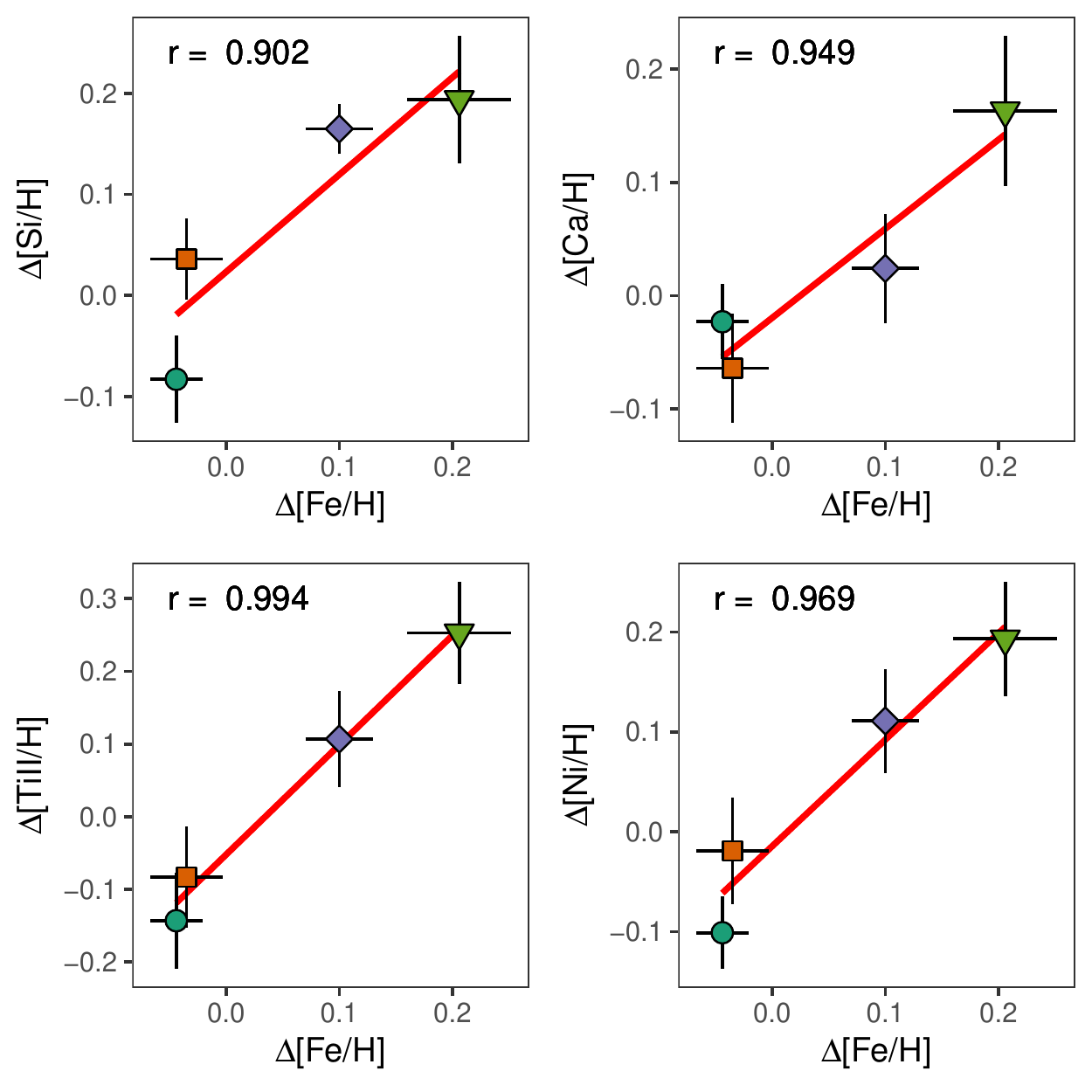}
\caption{Differential abundances (with respect S55974) for Si, Ca, Ti II, and Ni respectively, plotted against $\Delta$[Fe/H] for the analysed stars. Symbols are the same as in Fig.~\ref{fig:iron_spread}. The solid red line is a linear fit to the data. The Pearson correlation coefficient is reported in the left top corner.
\label{fig:corr_plots}}
\end{figure}


Figure~\ref{fig:iron_spread} shows the differential iron abundances $\Delta$[Fe/H] against the $\Delta_{\rm (F275W-F814W)}$ colour spread of target stars.  Differential iron abundances are highly correlated with the $\Delta_{\rm F275W-F814W}$ colour spread, as shown in Figure~\ref{fig:iron_spread}, in the sense that stars located at the blue end of the extended P1 sequence show systematically lower Fe abundances with respect to stars around the (0,0) origin of the chromosome map (Figure~\ref{fig:cmd}). This, again, is exactly what expected if the morphology of the extended P1 sequence in the cluster chromosome map is due to metallicity variations \citep{marino:19,legnardi,lscb}.

Binary stars can also contribute in principle to the extent of the P1 sequence towards negative $\Delta_{\rm F275W-F814W}$ values \citep{martins, marino:19}. From our data we cannot determine whether our sample contains one or more binaries. However, even if this were the case, the main result of our study would remain unaffected, since the observed metallicity dispersion is driven by the two stars located in the origin of the chromosome map, where the contribution of binaries is negligible \citep[see Fig. 9 in][]{marino:19}.

Next, we consider the trends of the differential abundances for each one of the other measured elements against iron. Those are shown in Fig.~\ref{fig:corr_plots}, along with the linear least-squares fits to the data and the Pearson correlation coefficients.
We can see that the differential abundances for each element show a statistically significant correlation with the iron counterparts. 

In the following, we explore the possibility that the observed abundance variations and positive correlations plotted in Fig.~\ref{fig:corr_plots} are not reflecting a genuine metallicity and
abundance dispersion in the cluster, but rather they are due to {\em (i)} an incorrect choice of stellar parameters or {\em (ii)} intrinsic variations in helium \citep[e.g.,][]{yong13}. As for possibility {\em (i)}, we simply note 
that differential abundance variations are measured for elements (Fe, Si, Ca, Ti~II, Ni) covering a variety of ionisation potentials and ionisation states. There is no single change in T$_{\rm eff}$, $\log$ g or v$_{\rm t}$ that would remove the observed abundance correlations for all elements in any given star. Thus, the fact that observed abundance variations are due to systematic errors in the stellar parameters is very unlikely \citep{yong13}. As for point {\em (ii)}, we know that helium variations are also able to account for the extended morphology of the P1 sequence; instead of a decrease of metal content, also an increase of the initial helium abundance would move stars towards lower values of 
$\Delta_{\rm (F275W - F814W)}$ in the chromosome maps, as shown by 
\citet{milone:15} and \citet{lsb18}.
Then the observed dispersion in metal to hydrogen ratios could be caused by a change of  the initial helium abundance (denoted by the helium mass fraction $Y$) rather than the metal content to iron (denoted here by the metal mass fraction $Z$). In fact, at a fixed $Z$ a change of $Y$ will change the hydrogen mass fraction $X$, such that the metal-to-hydrogen ratio $Z/X$ will change, because of the constraint $X+Y+Z$=1.
This option can be however discarded because in this case we would see stars at lower $\Delta_{\rm (F275W - F814W)}$ values having higher metal to hydrogen ratios (in this scenario $Y$ must be higher for  these stars, hence $X$ lower and $Z/X$ higher), which is the opposite of what is observed (see Fig.~\ref{fig:iron_spread}).

\begin{figure}
\centering
\includegraphics[width=0.98\columnwidth]{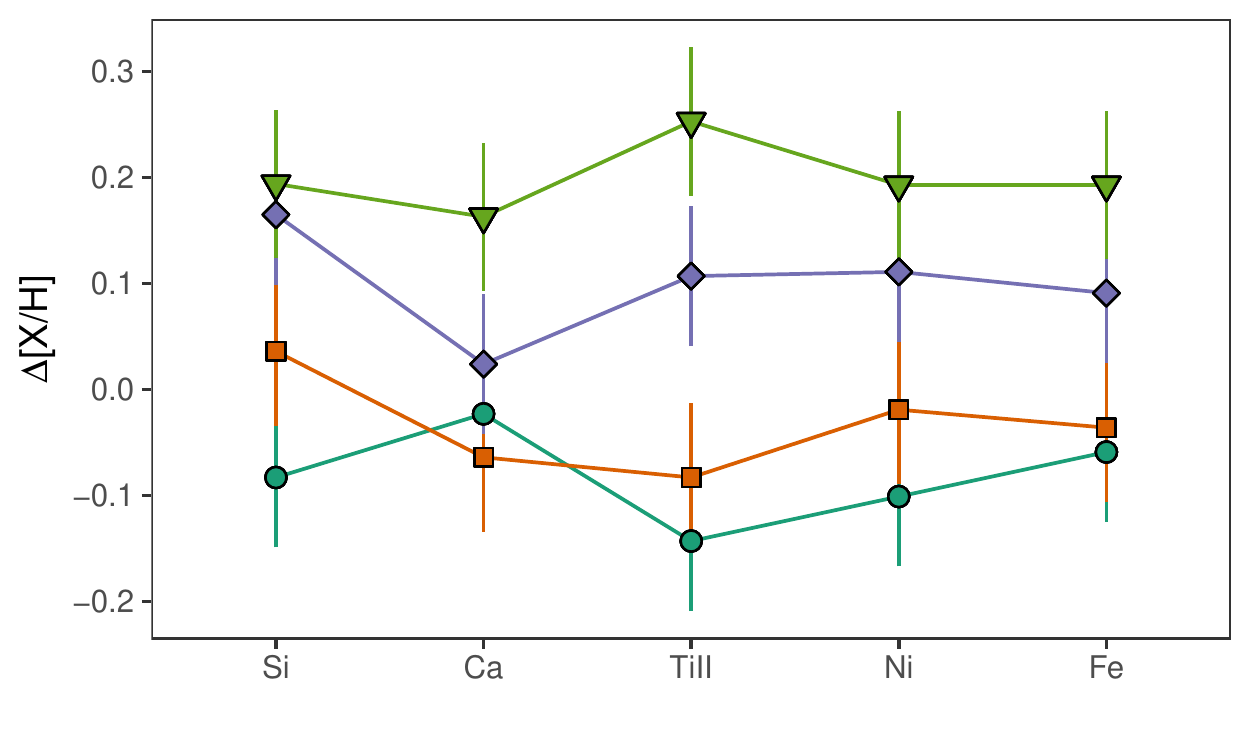}
\caption{Full set of differential abundances of Si, Ca, Ti II, Ni, and Fe for all the analysed stars. Symbols and colours are the same as in Fig.~\ref{fig:iron_spread}. Solid lines connect the set of abundances for each star. 
\label{fig:box_plot}}
\end{figure}

We show in Fig.~\ref{fig:box_plot} the full set of 
differential abundances of all measured elements for 
our sample of stars. As seen from the figure, the total ranges of differential 
abundances for the $\alpha$-elements Si, Ca, and Ti and the iron-peak element Ni are consistent, within errors, with the results for Fe.
This means the all these elements display comparable abundance spreads  among our P1 sample.
 {\em This would suggest that it is indeed the global metallicity that varies along the extended P1 sequence of this cluster}, and that a chemical enrichment by supernovae (most likely type II because of the $\alpha$-enhanced metal distribution of the reference star -- as shown in Table~\ref{table:par}) may be responsible for this phenomenon.


\section{Summary and Conclusions}
\label{conclusions}
We have performed a differential line-by-line analysis of five bright giants in NGC~2808 originally presented in \citet{cabrera:19}. Target objects are all members of the P1 group and were selected to have similar optical colours and magnitudes. We have obtained differential atmospheric parameters (e.g. with respect to a  reference star with similar parameters) for all stars in our sample using the standard excitation/ionisation balance technique and computed differential chemical abundances for Fe, Si, Ca, Ti, and Ni. 
Our differential line-by-line analysis of high-resolution spectra 
has allowed us to achieve high precision measurements of differential abundances, with average uncertainties for a given element as low as $\sim$0.03 dex. 

We have found that the range of differential abundances for all elements investigated are considerably larger than the average associated uncertainties, denoting 
the presence of intrinsic abundance spreads among our sample.
The total range of Fe abundance is equal to 
0.25$\pm$0.06~dex, with the lower Fe stars located at lower values 
of $\Delta_{\rm (F275W - F814W)}$ in the chromosome map, and 
higher Fe stars at higher $\Delta_{\rm (F275W - F814W)}$ values, 
as expected from photometry.
There are positive and statistically significant correlations between the values of the differential abundances for any given element and those for Fe, and the total ranges of differential abundances 
(hence the intrinsic abundance spreads in the sample) are all 
consistent, within errors, among the elements investigated.

Simulations by \citet{feng:14} show that the  chemical homogeneity of stars in a cluster is the result of turbulent mixing in the star-formation cloud gas -- e.g. the scatter in stellar abundances is at least five times smaller than that observed in the gas \citep[$\approx$ 0.06-0.3 dex over size scales of $\sim$0.1-1 kpc; e.g.  ][]{rosolowsky:08,bresolin:11,sanders:12}. Moreover, the process of star formation leads to a great amount of mixing, as soon as even very modest star formation efficiencies are attained. Thus, it is unlikely that the observed dispersion in metallicity of P1 stars is caused by internal variations within the gas out of which the star-forming cloud formed (unless different diffusion efficiencies during the cloud collapse are assumed; see also \citealp{legnardi}).

However, if a supernova occurs inside a star-forming cloud which was nearly homogeneous at the onset of the star formation and that continues forming stars thereafter, the change in iron abundance is expected to be measurable \citep[e.g.][]{bland:10}.
For example, in the theoretical model proposed by \citet{bailin:18}, giant molecular clouds (GMCs) can fragment into distinct clumps that undergo star formation at slightly different times. In such a scenario, core collapse supernovae from earlier-forming clumps can enrich clumps that have not yet begun forming stars, to the degree that the ejecta can be retained within the cloud potential well. This process then continues until these semi-independent clumps merge together to form the cluster. 

Along the same lines,~\citet{mckenzie21} find --from their hydrodynamical simulations of GMC formation in a high redshift dwarf galaxy-- that short-lived massive stars may increase the metallicity dispersion in a star forming GMC with initial homogeneous composition.
The merging of gas clumps and self-enrichment processes result in a metallicity dispersions of GC forming clumps of $\approx$0.1 dex, which may well explain the observed abundance variations in the P1 group, even if the exact amplitude of such variations largely depends on the initial metallicity and its radial gradient across the galaxy, the threshold gas density for star formation, and the star formation prescription. 

Our study confirms the presence of a metallicity spread amongst P1 stars in clusters with an extended P1 in their chromosome maps, as derived from photometry \citep{legnardi,lscb} and from the direct spectroscopic analysis of P1 stars in NGC~3201 \citep{marino:19}. The metallicity dispersions inferred from photometry by \citet{legnardi} for their sample of 55 Galactic GCs have an amplitude that is generally smaller than 0.15 dex. Systematic uncertainties in an absolute abundance analysis provide errors for individual measurements that are often comparable to, or larger than, the intrinsic variations of clusters themselves. Thus, the results presented here suggest that a differential abundance analysis can be best suited for any spectroscopic study at high-resolution, allowing to highlight even extremely small abundance variations that help understanding the mechanisms of formation of the GC multiple stellar  populations \citep{yong13,mckenzie22}.

\begin{acknowledgements}
M. Salaris acknowledges support from The Science and Technology Facilities Council Consolidated Grant ST/V00087X/1.
C. Lardo acknowledges funding from Ministero dell'Università e della Ricerca through the Programme {\em Rita Levi Montalcini} (grant PGR18YRML1).
Photometry used in this study is available at \url{https://archive.stsci.edu/prepds/hugs/}.
\end{acknowledgements}

%
\bibliographystyle{aa} 
\bibliography{biblio} 

\end{document}